\documentclass[amsmath,amssymb,twocolumn,prx,superscriptaddress]{revtex4-1}
\usepackage{graphicx}
\usepackage{dcolumn}
\usepackage{subfigure}
\usepackage{bm}
\usepackage{color}
\usepackage{braket}
\usepackage{hyperref}
\usepackage{amssymb}

\begin{document}

\title{Curvature screening in draped mechanical metamaterial sheets}
\author{Sourav Roy}
\affiliation{Department of Physics, Syracuse University, Syracuse, NY,  13244 USA}
\author{C.D. Santangelo}
\affiliation{Department of Physics, Syracuse University, Syracuse, NY,  13244 USA}

\date{\today}

\begin{abstract}
We develop a framework to understand the mechanics of metamaterial sheets on curved surfaces. Here we have constructed a continuum elastic theory of mechanical metamaterials by introducing an auxiliary, scalar gauge-like field that absorbs the strain along the soft mode and projects out the stiff ones. We propose a general form of the elastic energy of a mechanism based metamaterial sheet and specialize to the cases of dilational metamaterials and shear metamaterials conforming to positively and negatively curved substrates in the F\"{o}ppl-Von K\'{a}rm\'{a}n limit of small strains. We perform numerical simulations of these systems and obtain good agreement with our analytical predictions. This work provides a framework that can be easily extended to explore non-linear soft modes in metamaterial elasticity in future.  
\end{abstract}

\maketitle

\section{\label{sec:level1}Introduction}
The geometrical structure of mechanical metamaterials endow them with effective mechanical properties that can differ greatly from the materials from which they are fabricated \cite{lakes87, cherkaev1995elasticity, kolken2017auxetic}. A paradigmatic example can be constructed from counter-rotating, elastic polygons joined at nearly freely-rotating corners, which exhibits either a negative or positive Poisson ratio that is determined by the polygon geometry \cite{kagome, grima2013smart, Czajkowski2022, tobasco}. Understanding and characterizing the mechanical response of such materials has important applications in achieving advanced functionalities via topological protection, geometric frustration, non-linear responses and so on \cite{surjadi2019mechanical, bertoldi2017flexible}.

Here we consider mechanical metamaterials whose elastic properties arise from a single, global soft mode. For sufficiently large structures, however, generic arguments suggest a suite of additional modes with small elastic energies \cite{Czajkowski2022, tobasco}. These deformations look locally like the global soft mode of the metamaterial and dominate much of the global response of the structures to inhomogeneous forces. One might think of this in analogy to the  Nambu-Goldstone modes in thermodynamic systems and gauge theories with global symmetries \cite{nambu1960quasi, goldstone1961field}. For example, systems with a global, isotropic dilational mode exhibit universal deformations that approximate conformal transformations \cite{Czajkowski2022, tobasco} but there is still little known about how other metamaterial designs respond.

\begin{figure}[t]
    \includegraphics[width=0.48\textwidth]{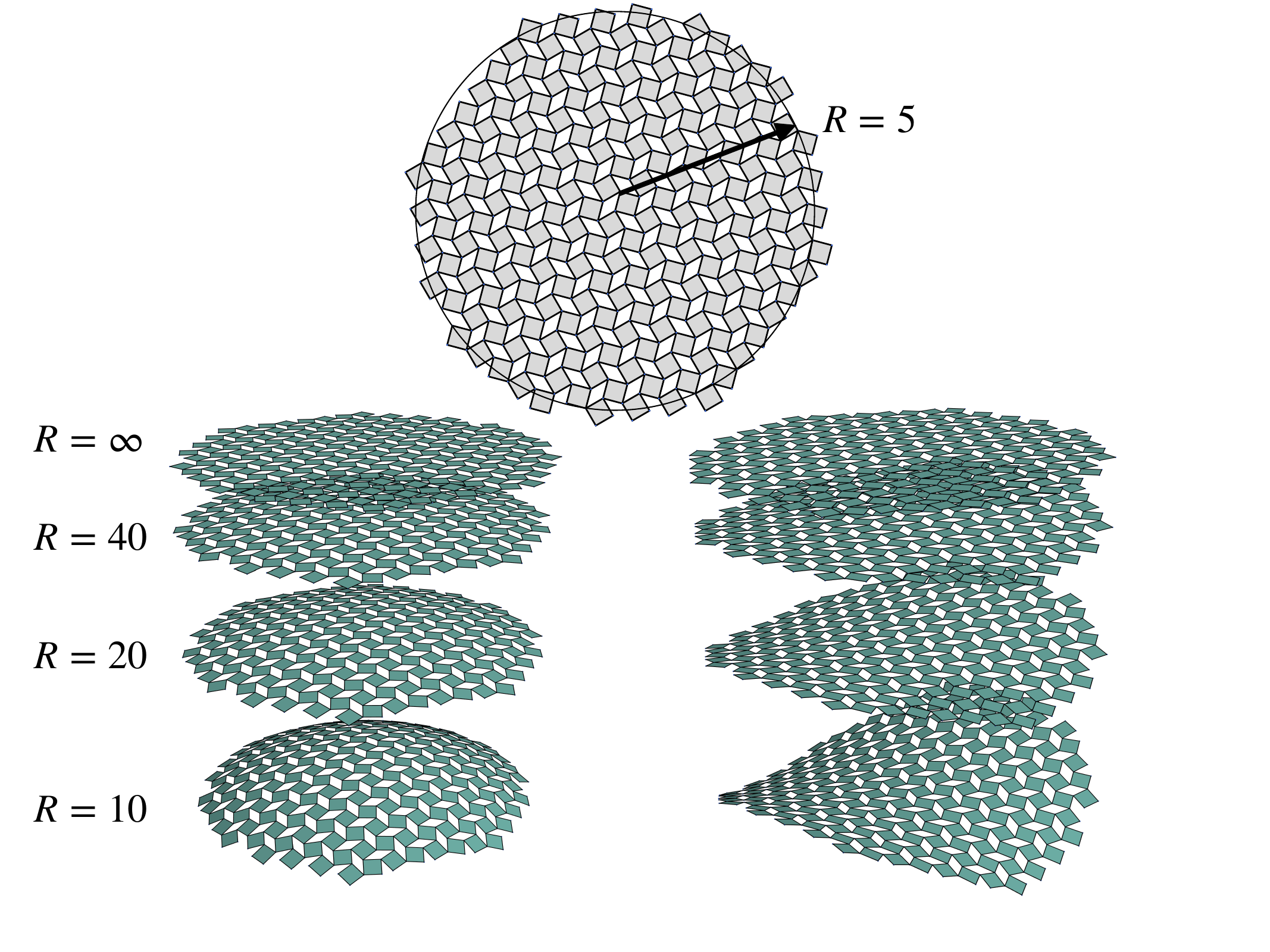}
    \caption{\label{fig1} Example of wrapping a conformal metamaterial onto a sphere or saddle of varying radii of curvatures. The internal displacements of the squares  give rise to a screening effect}
\end{figure}

One of the novel features of mechanical metamaterial sheets comes from their properties under bending \cite{konakovic2018rapid, jiang2022bending, nassar2022strain, konakovic2016beyond, greenwood2019developable, baek2018form}. When confining an elastic plate to a curved surface, the Gaussian curvature of the underlying substrate induces inhomogeneous stresses in the bulk of the sheet which is governed by the interplay between elasticity and geometry through Gauss' \textit{theorema egregium}.\cite{docarmodifferential} In a metamaterial, some of these stresses can be partially absorbed by the in-plane soft modes of the system (for example, Fig \ref{fig1} shows the soft in-plane displacements). Indeed, one expects that Gaussian curvature should be screened, much in the same way that disclination densities screen Gaussian curvature in curved crystalline membranes \cite{Seung, davidovitch2019geometrically}.

To better describe how geometry and elasticity of a metamaterial interact, we develop an approach to understand the low energy excitations of mechanical metamaterials confined to rigid, curved surfaces. We assume that the confinement is perfect but that the metamaterial is free to slide along the surface and that the sheet's boundaries are free. Instead of starting from a microscopic description of a particular metamaterial architecture, we employ a continuum description in terms of a prescribed metric, and introduce an auxiliary field to describe the soft mode. As a result, we obtain an effective elastic model describing the soft deformations of a metamaterial sheet that is easily applied to curved geometries.

We apply this approach to the deformations of dilational metamaterials as well as shear metamaterials which are metamaterials with zero energy uniform shear modes while they are kept confined to surfaces with positive and negative Gaussian curvature. Although we focus on structures that are well-approximated by Hookean elasticity, we will discuss how to generalize our methods to encompass nonlinear soft modes.

\section{The elastic energy of a metamaterial sheet}

\subsection{Mathematical formulation}
Our approach to the elasticity of metamaterial sheets is rooted in the relationship between global symmetries and their associated Nambu-Goldstone modes. Recall that a Nambu-Goldstone mode arises when a global symmetry, parameterized by a constant $\phi$ for example, is lifted to a slowly-varying, inhomogeneous field. Deformations that can be absorbed by changing $\phi$ have a particularly small energy cost that scales with gradients of $\phi$. And that energy cost can be made arbitrarily small by making $\phi$ vary slowly. For example, $\phi$ could measure the angle between two polygonal elements in a metamaterial that meet at a vertex, or represent a degree of freedom . Thus, we generically expect these soft deformation modes to dominate the elastic response of these materials.
This is precisely the approach taken by Zheng \textit{et al.} \cite{plucinsky, tobasco} to explore the soft deformations of mechanical metamaterials in 2D.

We represent the periodic metamaterial as a smooth surface indexed by coordinates $(x^1, x^2)$ with 3D positions $\mathbf{R}(x^1,x^2)$. The induced metric of the surface, $g_{i j}(x^1,x^2) = \partial_i \mathbf{R} \cdot \partial_j \mathbf{R}$, where $\partial_i = \partial/\partial x^i$, determines the local distances and relative angles between adjacent unit cells. The precise changes in the effective geometry of the metamaterial sheet are then encoded by a family of prescribed metrics, $\bar{g}_{i j}(\phi)$, which we also use to raise and lower indices \cite{sharon2010mechanics}. 

For a given, fixed value of $\phi$, the elastic strain, $\gamma_{i j} = [g_{i j} - \bar{g}_{i j}(\phi)]/2$, measures how the sheet deforms relative to its prescribed metric \cite{sharon2010mechanics}. This form of the strain contains a component that can be directly absorbed by a spatially-dependent change in $\phi$ -- the metamaterial soft mode -- and an orthogonal component that cannot. To extract the portion associated with changes along the soft mode, we first consider a small change in the parameter $\phi$ to $\phi + \delta \phi$, which leads to a change in the prescribed metric, $G_{i j} \equiv \partial_\phi \bar{g}_{i j}(\phi)$. Thus, any strain can be written as $\gamma_{i j} = \gamma ~ G_{i j} + \gamma^\perp_{i j}$, where $\gamma^\perp_{i j}$ is a part of the strain associated to deformations orthogonal to the soft modes  and $\gamma$ is $\propto$ $\delta \phi$. The scalar part in the first term, $\gamma = H^{i j} \gamma_{i j}$ is the ``soft'' strain where, $H^{i j} = G^{i j}/G^{k l} G_{k l}$  and $H^{i j} \gamma^\perp_{i j} = 0$.

More generally, we can also consider metamaterials with local prescribed curvature tensor $\bar{h}_{i j}(\phi)$. This can arise, for example, when the global soft mode of the metamaterial acts to change its curvature as well as the in-plane geometry. In this paper, we will assume that $\bar{h}_{i j}(\phi) = 0$ and, in fact, that the thickness of our material is sufficiently small that bending stresses can be entirely neglected compared to in-plane stresses\cite{landau1986theory}. This  allows us to isolate the behavior we want to study, how the internal soft mode responds to Gaussian curvature, without additional complications.

Since the deformation represented by $\gamma$ and that represented by spatially varying $\delta \phi$ are redundant, we make the essentially arbitrary choice to fix $\phi$ to a given constant value and retain all deformations in terms of the strain. A general elastic energy can be written as
\begin{equation}\label{eq:E1}
    E_1 = \int dS ~ \mathcal{W}(\gamma , \partial_i \gamma , \gamma_{i j}^\perp),
\end{equation}
where $dA$ is the area measure determined by the prescribed metric. Note that, while the energy density, $\mathcal{W}$, depends explicitly on the portion of the strain directed along the metamaterial soft mode, the assumed softness of the global mode suggests including a dependence on the gradient as well.

\subsection{An auxiliary field represents the soft mode}
In principle, we can proceed directly with Eq. (\ref{eq:E1}). However, working with the orthogonal projection of the strain, $\gamma_{i j}^\perp$, proves challenging. It can be avoided if we introduce an auxiliary field, $A$. We then consider the elastic energy
\begin{equation}\label{eq:E2}
    E_2 = \int dS \left[ \mathcal{W}(\gamma , \partial_i \gamma , \gamma_{i j} - A G_{i j}) + \frac{1}{2} c_A (\gamma - A)^2 \right].
\end{equation}
To understand the role of the auxiliary field $A$, consider the simultaneous transformations $\gamma_{i j} \rightarrow \gamma_{i j} + \delta \phi G_{i j}$ and $A \rightarrow A+ \delta \phi$, which leave $\gamma_{i j} - A G_{i j}$ and $\gamma - A$ invariant. This ensures that $E_2$ transforms identically to $E_1$ and, as we will see below, results in the same equilibrium.

To demonstrate their equivalence, we introduce the following notations for derivatives of $\mathcal{W}$: $\partial \mathcal{W}$ indicates a derivative with respect to the first argument, $\partial^i \mathcal{W}$ is a derivative with respect to the vector-valued second argument, and $\partial^{i j} \mathcal{W}$ is a derivative with respect to the tensor valued third argument. Then the equilibrium equation for the field $A$  reads
\begin{equation}\label{eq:equil1}
    -G_{i j} \partial^{i j} \mathcal{W}(\gamma , \partial_i \gamma , \gamma_{i j} - A G_{i j}) + c_A (A - \gamma) = 0.
\end{equation}

The second derivative of the energy with respect to $A$ is $G_{i j} G_{k l} \partial^{i j} \partial^{k l} \mathcal{W} + c_A$, where the arguments of $\mathcal{W}$ have been suppressed. This is positive-definite if $c_A$ is sufficiently large.
While the precise value of $c_A$ is irrelevant to the physics, its largeness ensures that the solution $A$ to Eq. (\ref{eq:equil1}) gives the minimum of the energy.

We compute the stress tensor next by varying $E_2$ with respect to $\gamma_{i j}$ and substituting the result of Eq. (\ref{eq:equil1}). We obtain
\begin{equation}
    \sigma^{i j} = \partial^{i j} \mathcal{W} - H^{i j} G_{k l} \partial^{kl} \mathcal{W} + H^{i j} \left( \partial \mathcal{W} - D_k \partial^k \mathcal{W} \right),
\end{equation}
where $D_k$ is the covariant derivative with respect to the prescribed metric. This is precisely what one obtains directly from $E_1$, showing that $E_1$ and $E_2$ are equivalent.

Finally, we consider further simplification of the elastic energy,
\begin{equation}\label{eq:E3}
    E_3 = \int dS ~\left[ \mathcal{W}(A, \partial_i A , \gamma_{i j} - A G_{i j}) + \frac{1}{2} c_A (\gamma - A)^2 \right]
\end{equation}
in which we have replaced almost every instance of $\gamma$ with $A$. A similar model was introduced in Ref. \cite{kagome} to describe the twisted, kagome lattice in two dimensions. As we will see, Eq. (\ref{eq:E3}) will allow us to apply the F\"oppl-von K\`arm\`an limit to a curved metamaterial in a fairly straightforward way.

However, it turns out that $E_3$ is not exactly equivalent to $E_1$ and $E_2$. We can see this from the equilibrium equation for $A$,
\begin{eqnarray}\label{eq:recursion}
    A &=& \gamma - \frac{1}{c_A} \left[\partial \mathcal{W} - D_i \partial_i \mathcal{W} - G_{i j} \partial^{i j} \mathcal{W} \right].
\end{eqnarray}
For large $c_A$, Eq. (\ref{eq:recursion}) can be interpreted as a recursion relation from which one obtains $A$ in terms of $\gamma$ and its derivatives as a series in $1/c_A$. Eq. (\ref{eq:recursion}), becomes asymptotically exact as $c_A \rightarrow \infty$ and the stress tensor of $E_3$ approximates that of $E_1$. Since physical quantities do not depend on the precise value of $c_A$ and since the limit $c_A \rightarrow \infty$ leads to analytically tractable regimes, we will adopt this limit in the remainder of the paper.

Our approach has been quite general up to this point, but now we expand the energy density $\mathcal{W}$ assuming that $\gamma_{i j} - A G_{i j}$ and $\partial_i A$ are small. This leaves
\begin{eqnarray}\label{eq:W}
    \mathcal{W} &\approx& W_0(A) + \frac{1}{2} \kappa_1 \bar{g}^{i j} \partial_i A \partial_j A \\
\nonumber       &  &+ E^{i j k l} (\gamma_{i j } - A G_{ i j}) (\gamma_{k l } - A G_{ k l}),
\end{eqnarray}
where the elastic tensor $E^{ i j k l}$ is assumed to have no dependence on $A$ at lowest order. For a metamaterial with a global zero mode, $W_0(A) = 0$, though more generally, we may expect $W_0(A)$ to be a nonlinear function of $A$. In what follows, however, we will assume that $W_0(A) = \kappa_0 A^2/2$ where $\kappa_0$ is small.
\begin{figure}[t]
    \includegraphics[width=0.48\textwidth]{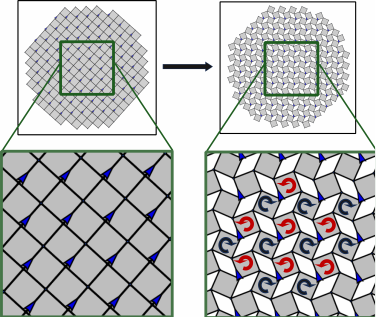}
    \caption{\label{fig2} The undeformed canonical square metamaterial (left), upon extensive loading, deforms by counter-rotation of the squares (right). The red and blue arrows show the direction of rotation of adjacent squares as the metamaterial dilates.}
\end{figure}

\section{Examples and Results}

\subsection{Dilational Metamaterial}
To see how Eq. (\ref{eq:E3}) works in practice, consider a 2D, isotropic dilational material, for which $G_{i j} = \bar{g}_{i j}$. We obtain an elastic energy of the form
\begin{eqnarray}\nonumber
   E    &\approx& \int dS ~ \left[ \frac{1}{2} E^{i j k l} (\gamma_{ i j } - A \bar{g}_{i j} ) (\gamma_{ k l } - A \bar{g}_{k l} )  \right. \\
        & & \left. + \frac{\kappa_0}{2} A^2 + \frac{\kappa_1}{2} \bar{g}^{i j} \partial_i A \partial_j A \right].
\end{eqnarray}
where $E^{i j k l} = c_A \bar{g}^{i j} \bar{g}^{k l} + \mu (\bar{g}^{i k} \bar{g}^{j l} - \bar{g}^{i j} \bar{g}^{k l})$ and $\mu$ is the shear modulus of the material. Varying this elastic energy with respect to $A$ yields
\begin{equation}
    \gamma = \left(1 + \frac{\kappa_0}{c_A} \right) A - \frac{\kappa_1}{c_A} \triangle A
\end{equation}
and can be inverted to
\begin{equation}\label{eq:A}
   A = \frac{c_A}{c_A + \kappa_0} \gamma + \frac{\kappa_1 c_A}{(c_A + \kappa_0)^2} \triangle \gamma + \mathcal{O}(\triangle^2 \gamma).
\end{equation}
As expected, the limit $c_A \rightarrow \infty$ implies $A \rightarrow \gamma$. The stress tensor is
\begin{equation}
    \sigma^{i j} = (\kappa_0 A - \kappa_1 \triangle A) \bar{g}^{i j} + E^{i j k l} (\gamma_{k l} - A \bar{g}_{k l}).
\end{equation}
Using Eq. (\ref{eq:A}), we obtain
\begin{equation}\label{eq:sigma_conformal}
    \sigma^{ i j } = \left( c_0 \gamma - c_1 \triangle \gamma \right) \bar{g}^{ i j } + E^{i j k l} \gamma_{k l},
\end{equation}
where
\begin{eqnarray}
    c_0 &=& \frac{ \kappa_0 c_A }{c_A + \kappa_0} \xrightarrow{c_A \rightarrow \infty} \kappa_0\\
    c_1 &=& \frac{ \kappa_1 c_A^2}{(c_A + \kappa_0)^2} \xrightarrow{c_A \rightarrow \infty} \kappa_1. \nonumber
\end{eqnarray}

To understand how such a sheet conforms to a weakly curved surface, we now specialize to the F\"{o}ppl-Von K\'{a}rm\'{a}n limit. Displacements in the $xy-$plane are given by a vector $u^i$ and displacements in $z$ are given by a function $\zeta$, so that $\gamma_{i j} \approx (\partial_i u_j + \partial_j u_i)/2 + \partial_i \zeta \partial_j \zeta/2$.
In this limit, we write $\sigma^{i j} = \epsilon^{i k} \epsilon^{j l} \partial_k \partial_l \chi$ in terms of an Airy potential $\chi$. From Eq. (\ref{eq:sigma_conformal}), we obtain
\begin{eqnarray} \label{eq:eqnofmotion1}
    \triangle \chi    &=& \kappa_0 A - \kappa_1 \triangle A
\end{eqnarray}
In addition to this, the geometric compatibility equation, required for $\chi$ to be expressed in terms of a displacement $u^i$, is
\begin{equation}\label{eq:eqnofmotion2}
    \frac{1}{Y} \triangle^2 \chi = - G - \Delta A,
\end{equation}
where $G$ is the local Gaussian curvature and the effective Young's modulus for the metamaterial sheet is $Y = 4 \mu (c_A + \mu)/(2 \mu + c_A)$,  where $Y \rightarrow 4 \mu$ as $c_A \rightarrow \infty$.

Eq. (\ref{eq:eqnofmotion2}) shows that the motion along the soft mode of the metamaterial screens the elastic stresses induced by the Gaussian curvature, in a manner similar to how disclinations in a 2D crystal screen Gaussian curvature. We can better understand the form of Eq. (\ref{eq:eqnofmotion2}) by considering an isothermal coordinate system on a curved surface.Recall that, in an isothermal coordinate system, the metric of a surface is proportional to the Euclidean metric, i.e., $g_{i j} = (1 + 2 ~\delta \Omega(x,y)) \delta_{i j}$, where $(1+ 2\delta \Omega(x,y))$ is the local conformal factor \cite{docarmodifferential}.In such a coordinate system, the Gaussian curvature takes a particularly simple form, $G = - \triangle \delta \Omega$ to first order in $\delta \Omega$. Thus, it appears that the auxiliary field $A$ plays precisely the role of $\delta \Omega$ in the conformal factor, as one might expect in a dilational metamaterial.

To understand Gaussian curvature screening, we can eliminate $\chi$ to obtain the screened Poisson equation,
\begin{equation} \label{eq:auxiliaryfieldeq1}
    \left[\frac{\kappa_1}{4\mu} \triangle - \left( \frac{\kappa_0}{4\mu}+ 1\right) \right] \triangle A = G.
\end{equation}
where $G$ acts as a source and the screening length is $l_{sc} = \sqrt{\kappa_1/(4 \mu + \kappa_0)}$.
Equivalently, one can eliminate $A$ and obtain a screening law for $\chi$ by introducing a geometric potential $\Omega$ \cite{vitelli2006crystallography} such that $\triangle \Omega = G$. Then one obtains
\begin{equation}
    \frac{1}{4\mu}\triangle \chi = -\Omega - A + h_1
\end{equation}
where $h_1$ is a harmonic function determined by boundary conditions on $\chi$, $A$, and the geometric potential.
Thus, we find
\begin{equation}\label{eq: screeninginstress}
    \left[\frac{\kappa_1}{4\mu} \triangle - \left( \frac{\kappa_0}{4\mu}+ 1\right) \right] \triangle \chi = -\kappa_1 G +\kappa_0 \Omega -\kappa_0 h_1.
\end{equation}
so that the dilational stress, $\triangle \chi$, satisfies the screened Poisson equation with a source $\kappa_1 G - \kappa_0 \Omega + \kappa_0 h_1$ and screening length $l_{sc}$. A key point to note here is the contrast between Eq. \ref{eq: screeninginstress} and that of a Hookean elastic sheet where we have simply $\triangle^2 \chi \propto G$. In the case of Eq. (\ref{eq: screeninginstress}) stresses are relaxed by the soft mode over a length scale $l_{sc}$, in marked contrast to a Hookean elastic sheet.
\begin{figure}[t]
    \includegraphics[width=0.48\textwidth]{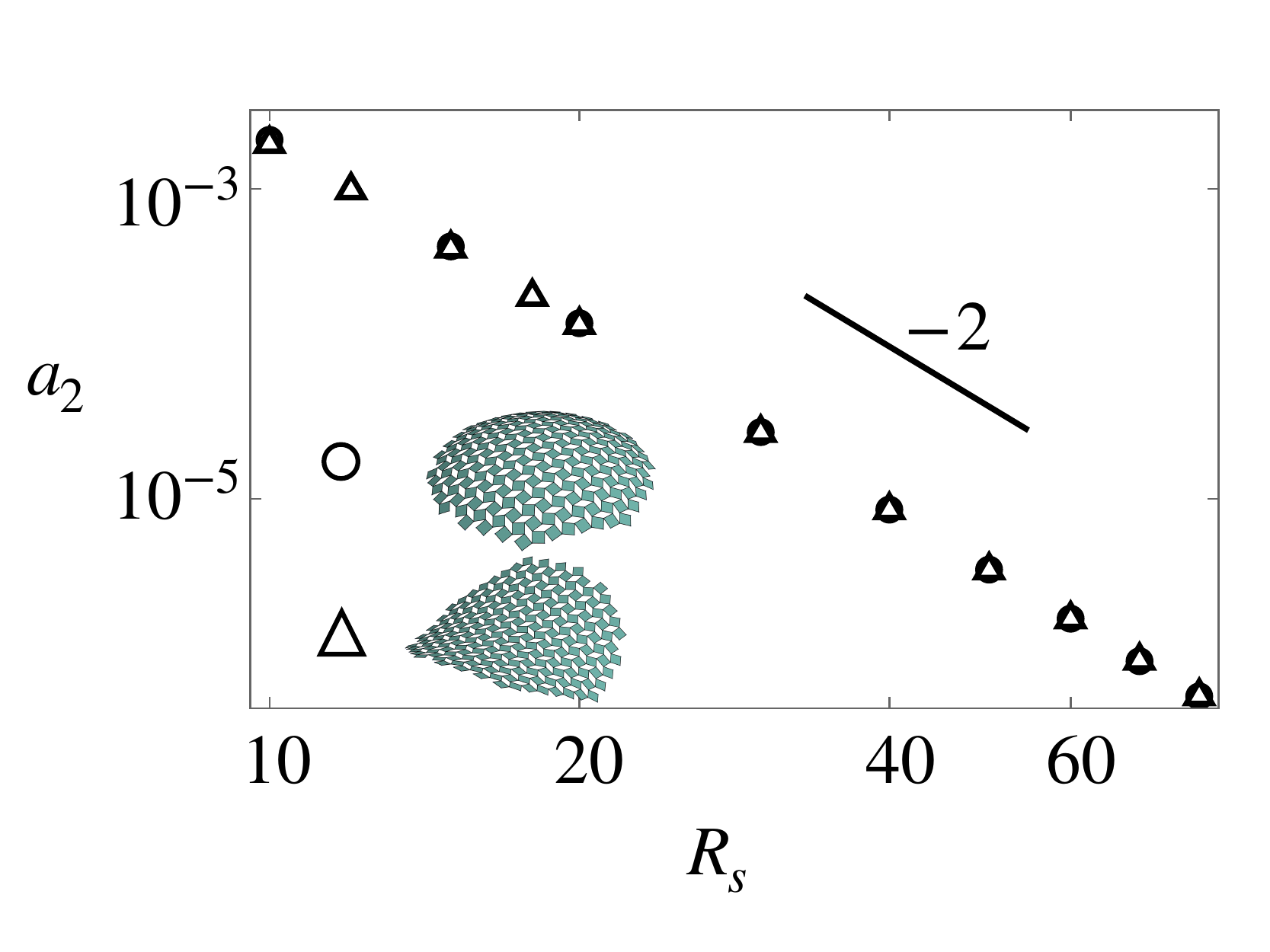}
    \caption{\label{fig3} Coefficient $a_2$ found by fitting the conformal factor, $A = a_0 + a_2 r^2$ as a function of radius of curvature $R_s$ is consistent with $a_2 \propto R_s^{-2}\propto |G|$,} from the analytical prediction.
\end{figure}
\begin{figure}[t]
    \includegraphics[width=0.48\textwidth]{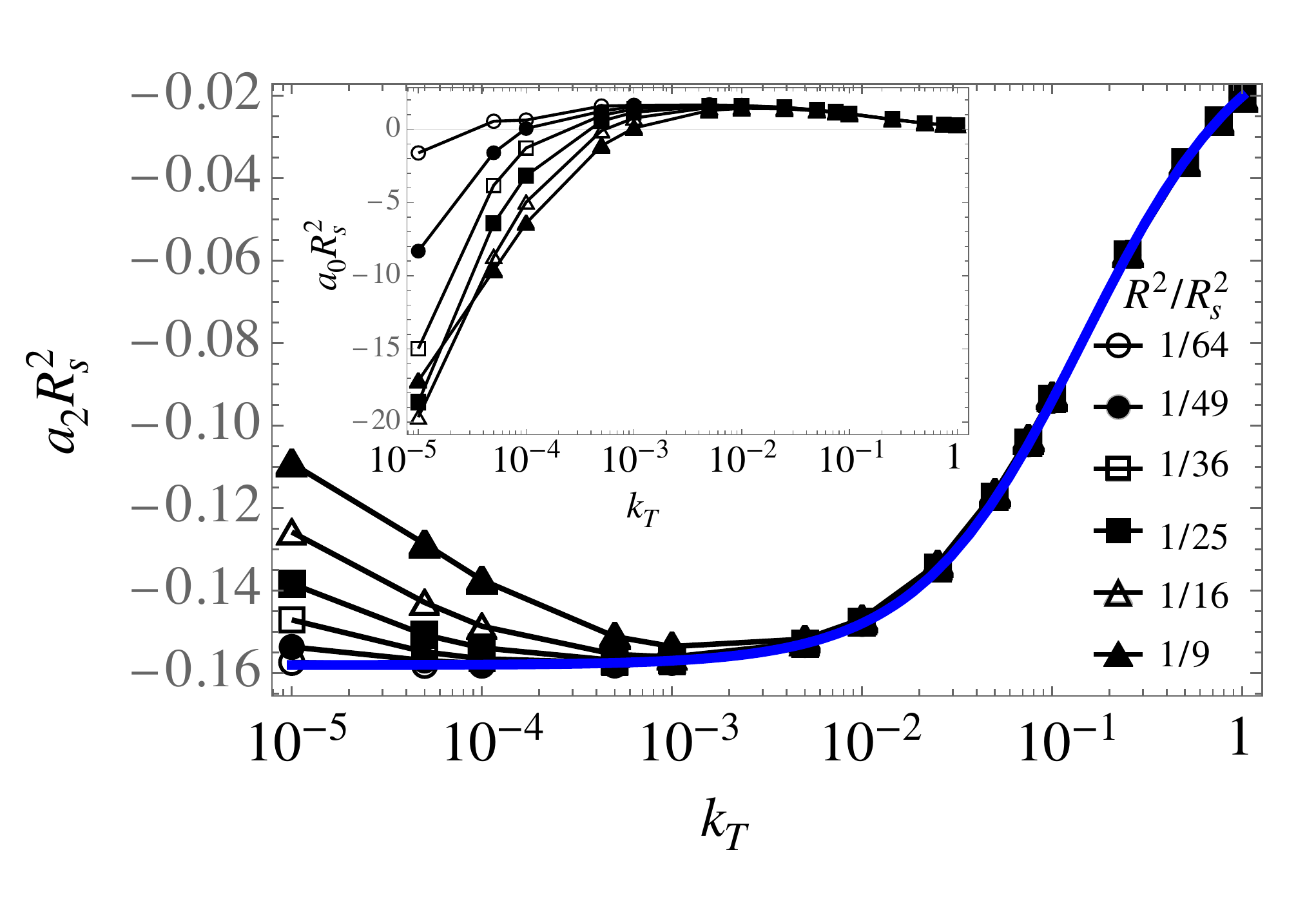}
    \caption{\label{fig4} $a_2 R_s^2$} as a function of torsional stiffness of rotating square metamaterial. The blue curve is a best fit of $b_1$ and $b_2$ to the theoretical prediction $a_2 R_s^2 \propto  b_1/(1 + b_2 k_T)$ for the largest radius sphere, where $k_T$ is the torsional spring constant. Inset shows curves of $a_0 R_s^2$
\end{figure}
To obtain an explicit solution, we assume the metamaterial is a disk of radius $R$ adsorbed to a surface of constant Gaussian curvature $G$. Then we obtain
\begin{eqnarray}
    A       &=& a_0 + a_2 r^2 + a_b I_0( r/l_{sc} ) \nonumber \\
    \chi    &=& d_0 + d_2 r^2 + d_4 r^4 + d_b I_0(r / l_{sc} ) \nonumber
\end{eqnarray}
subject to free boundary conditions  $\partial_r \chi|_{r=R} = 0$, and $\partial_r A|_{r=R} = 0$.  The choice of boundary conditions and constant Gaussian curvature G gives, $\Omega = \frac{G}{4} r^2$ and the $h_1$= constant. When $\kappa_0 \ne 0$, this is sufficient to determine all coefficients up to constant $d_0$ (SI).
In the limit $c_A \rightarrow \infty$, we find
\begin{eqnarray}
    A       &=& \frac{\mu G (R^2-12 l_{sc}^2)}{3 (4 \mu+\kappa_0)} - \frac{\mu G r^2}{(4 \mu + \kappa_0)} \\
\nonumber   & & + \frac{2 \mu G ~R l_{sc}}{(4 \mu+\kappa_0)} \frac{I_0(r/ l_{sc})}{I_1(R/l_{sc}))}.
\end{eqnarray}

To test the resulting prediction for $A$, we performed numerical simulation of a metamaterial built from counter-rotating squares shown in Fig.\ref{fig2} . The edges of each square are springs with spring constant $k_S$  (Section 2 in SI). To prevent the squares from bending, we add an additional hidden vertex above and below the centroid of each square and connect them to the square's vertices (these are sometimes called ``blocks''; see \cite{guest2018mobility, finbow2013isostatic}). The spring constant of the additional springs, $k_B \le k_S$, sets the bending stiffness of the squares.
We also incorporate a torsional spring on every other joint between adjacent squares. The elastic energy of the joint is $E_{j} = k_T (\theta - \pi/4)^2/2$, where $\theta$ is the angle between adjacent squares. On a flat surface, the equilibrium angle of the torsion joints, $\pi/4$, ensures that the sheet can express both compression and extension relative to when draped on a curved surface as necessary.  All the lengths in the simulation are defined relative to two times the square edge spring equilibrium length. Finally, we confine the vertices of the square panels to a surface of shape $x^2 + y^2+ (z-r)^2 = R_s^2$ (sphere) or $x^2 - y^2 + (z-r)^2 = R_s^2$ (saddle), but otherwise allow the vertices to slide along the surface.

\begin{figure}[t]
    \includegraphics[width=0.48\textwidth]{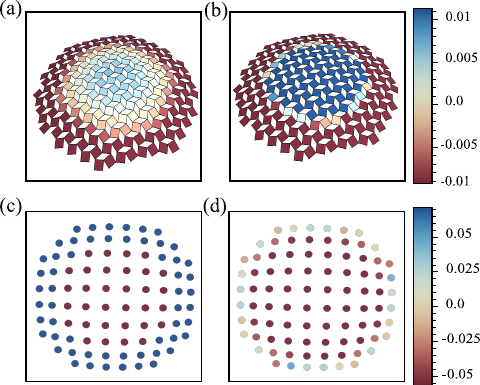}
    \caption{\label{fig5} The dilational area strain of the squares, $(\alpha - \alpha_0)/\alpha_0$, measured by comparing the area of each square, $\alpha$, to its equilibrium area $\alpha_0$ for (a) a floppy ($k_T = 0.01$) metamaterial and for (b) a stiff ($k_T= 10$) metamaterial. (c, d) The fractional angular deformation, $(\theta - \theta_0)/\theta_0$ of the torsional springs in both the floppy ($k_T = 0.01$) and stiff ($k_T= 10$) metamaterials, as seen from above. This shows the trade-off between curvature-induced stresses and the metamaterial soft mode. The area of the spherical surface is $R_s = 15$.}
\end{figure}

Fig. \ref{fig1} illustrates the model and the geometries formed when draping over both a spherical cap and a saddle.
The field $A$ is computed by directly mapping the spring angle $\theta$ to the corresponding unit cell size on a flat geometry.
We fit the measured $A$ to functional form $A = a_0 + a_2 r^2$ obtained from our analytical theory for both saddle and sphere geometries in Fig. \ref{fig3}, showing good agreement with the expected dependence of $A \propto G$. In Fig. \ref{fig4}, we plot 
$a_0 R_s^2$ and $a_2 R_s^2$ as a function of the torsional stiffness, $k_T$, on spheres with several values of radius $R$. At larger $k_T$ and smaller curvatures, we see that the corresponding curves collapse into a universal curve, though we note some deviations at smaller $k_T$ that depend on the sphere radius.  In Fig. \ref{fig5}, we plot the stress distribution along with the corresponding soft modes to verify the screening effect predicted in Eq. \ref{eq: screeninginstress}.

Generally, we expect that $\kappa_0$ is proportional to $k_T$ so we compute the best fit to the theoretical prediction  $a_2 R_s^2 = b_1/(1+ b_2 k_T)$ (blue curve) to the data from largest sphere. While the results of the simulations fit the theory at the smallest curvatures, as the curvature increases we see systematic deviations at small $k_T$. This may be explained by the failure of the F\"oppl von K\'arm\'an limit at high curvatures or the increased relative importance of bending energy in the simulation at the smallest values of $\kappa_0$.

\subsection{Shear metamaterials}
Here we study a simple shear mechanism based metamaterial. As a concrete example, we may consider a square grid of cross-linked, stiff fibers, where we assume the angle at which the fibers meet is flexible. The schematic for this is shown in \ref{fig4} (inset), where the material is wrapping curved substrates. These have been called elastic gridshells in literature and their non-trivial geometries upon buckling from a planar grid of fibers have been studied in detail\cite{baek2018form}, showing that the inextensibility of the cross-linked fibers play a key role in the final shapes of these materials.
In the continuum, for a simple shear mechanism, we have
\begin{equation}
    G_{i j} = \begin{pmatrix}
        0 & 1 \\
        1 & 0
    \end{pmatrix}.
\end{equation}
Minimizing the elastic energy with respect to $A$ yields
\begin{equation}
    -\frac{1}{2} G_{i j} \sigma^{i j} + \kappa_0 A - \kappa_1 \triangle A = 0.
\end{equation}
The corresponding geometrical equation can be found by noting that, in the F\"oppl von K\'arm\'an limit, $\epsilon^{i k} \epsilon^{j l} \partial_i \partial_j \gamma_{k l} = -G + \partial_x \partial_y A$. After some algebra, we obtain
\begin{equation}
    \frac{B+\mu}{4 \mu B} \triangle^2 \chi + \frac{\mu - c_A}{\mu c_A} (\partial_x \partial_y)^2 \chi = -G + \partial_x \partial_y A.
\end{equation}
where, B is the elastic bulk modulus. It is easy now to take $c_A \rightarrow \infty$ in Eq. (\ref{eq:sheareq}) to obtain
\begin{equation}\label{eq:sheareq}
    \frac{B+\mu}{4 \mu B} \triangle^2 \chi - \frac{1}{\mu} (\partial_x \partial_y)^2 \chi = -G + \partial_x \partial_y A.
\end{equation}
Eq.(\ref{eq:sheareq}) is similar in spirit to Eq.(\ref{eq:eqnofmotion2}) and thus captures how metamaterial soft modes screen out the stress sourced by the Gaussian curvature of the substrate. To understand the form of the auxiliary field entering the geometric equation, we can consider a coordinate system on the surface called a Tchebyshev net\cite{lirias1594633}, which has metric $ds^2 = du^2 + dv^2 + 2 \sin \omega(u,v) ~ du dv$. This can be thought of as precisely the deformation of inextensible but bendable fibers that meet at local angles, $\omega$. The Gaussian curvature is given by $G = -\partial_u \partial_v \omega/\sin \omega$. Writing $\omega = \pi/2 - \delta \theta$, we obtain $G \approx \partial_u \partial_v \delta \theta$. Thus, we can identify $A$ with the change in angle at which intersecting fibers meet.
\begin{figure}[t]
    \includegraphics[width=0.48\textwidth]{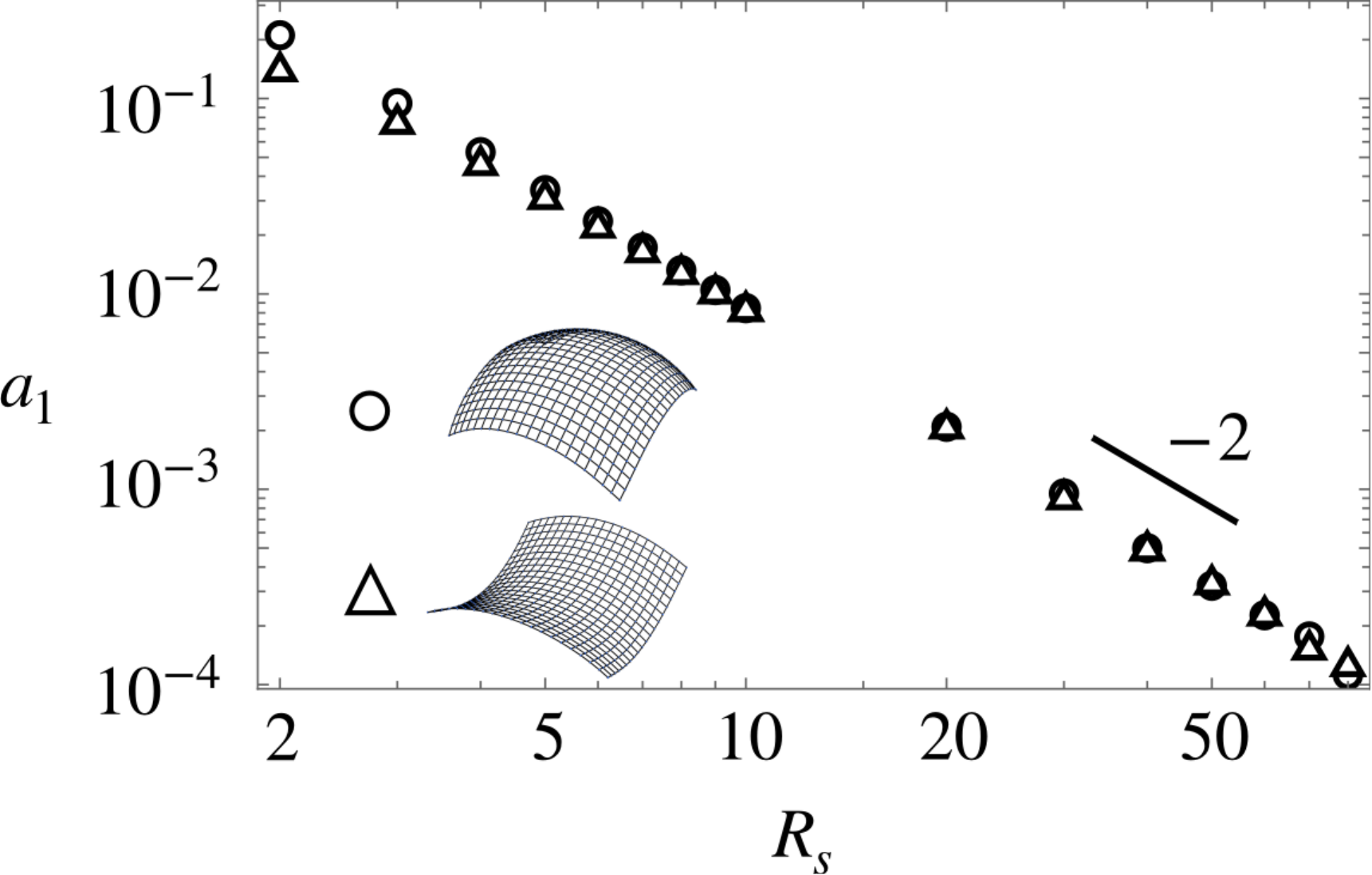}
    \caption{\label{fig6}  Simulations for $A$ at the center of a simple shear metamaterial on sphere (open circles) and saddle (triangles). The value of $|a_1|$ is determined by the best fit of the $\pi/2 - \theta$ to $a_0 + a_1 x y$. The results are consistent with the formula $|A| \propto R_s^{-2} \propto G$.}
\end{figure}
A similar analysis for a pure shear metamaterial instead yields the geometric equation,
\begin{equation}
    \frac{1}{4 B} \triangle^2 \chi + \frac{1}{\mu} (\partial_x \partial_y)^2 \chi = -G + \frac{1}{2} ( \partial_x^2 - \partial_y^2  ) A,
\end{equation}
and
\begin{equation}
    \frac{1}{2} ( \partial_x^2 - \partial_y^2  ) \chi + \kappa_0 A - \kappa_1 \triangle A = 0.
\end{equation}
Despite the superficial difference between simple and pure shear metamaterials, a straightforward calculation shows that these equations are identical provided one rotates the coordinate system by $\pi/4$, which recovers the relationship between simple and pure shear from the 2D elasticity of uniform media.

Eq. (\ref{eq:sheareq}) is difficult to solve since it mixes hyperbolic and elliptic operators in an unusual way.
However, for constant and small $G$, we obtain a solution when $\kappa_1$ is small by dividing the sheet into bulk and boundary parts. We replace $Y = \frac{4 \mu B}{B+\mu}$ in the equations below.
In the bulk of a rectangular domain where $x$ and $y$ both span from $-L$ to $L$, we look for a solution of the form $A_{bulk} = a_1 x y$ and $\chi = b_1 x^2 y^2$, finding
\begin{eqnarray}\nonumber
    a_1 &=& \frac{G Y \mu}{2 \kappa_0 \mu + Y (\mu - \kappa_0)} \\
\nonumber
    b_1 &=& -\frac{\kappa_0}{4} \frac{G Y \mu}{2 \kappa_0 \mu + Y (\mu - \kappa_0)}.
\end{eqnarray}

To match the boundary conditions, these solutions must be augmented by a boundary layer of width $l_{shear} = \sqrt{\kappa_1/\kappa_0}$. While finding the exact form of this boundary layer is difficult, to lowest order in $\kappa_1$ and $G \kappa_0$ we obtain an approximate solution,
\begin{eqnarray}
 \label{eq:shearA} 
    A       &=& A_{bulk} + A_{boundary} \\
\nonumber   &\approx&  \frac{\mu Y G }{   Y(\mu -\kappa_0) +2 \kappa_0 \mu  } \Big[ x y + \\
\nonumber   & & + l_{shear} \frac{  x~ \sinh(y/l_{shear} ) + y ~ \sinh(x/l_{shear}) }{\cosh \left( L/l_{shear} \right)} \Big]
\end{eqnarray}
and
\begin{eqnarray}
    \chi    &=& \chi_{bulk} + \chi_{boundary} \\
\nonumber   &\approx& \frac{\mu Y G/4}{Y(\mu -\kappa_0)+ 2\kappa_0 \mu }\Big[ L^2(x^2+y^2) -x^2 y^2 \\
\nonumber   & & - 4 Y l_{shear}^4~
   \frac{ \cosh(x/l_{shear})+\cosh(y/l_{shear})}{\cosh (L/l_{shear})}\Big].
\end{eqnarray}

We also perform numerical minimization of a shear metamaterial built from a square net of filaments. The bending resistance of the filaments are enforced by torsional springs with stiffness $k_b$, however we also include torsion springs with modulus $k \ll k_b$ on each square to resist the soft shear mode. Conjugate gradient energy minimization of this shape on both the sphere and saddle are performed and the internal mechanism is measured from the angle filaments when they cross. The resulting angle, measured with respect to the equilibrium angle $\pi/2$, is fit to $A = a_0 + a_1 x y$, which accurately reflects the measured values of $A$ away from the boundary. In Fig. \ref{fig6}, we see that $a_1 \propto |G|$ as expected.

\begin{figure}[t]
    \includegraphics[width=0.48\textwidth]{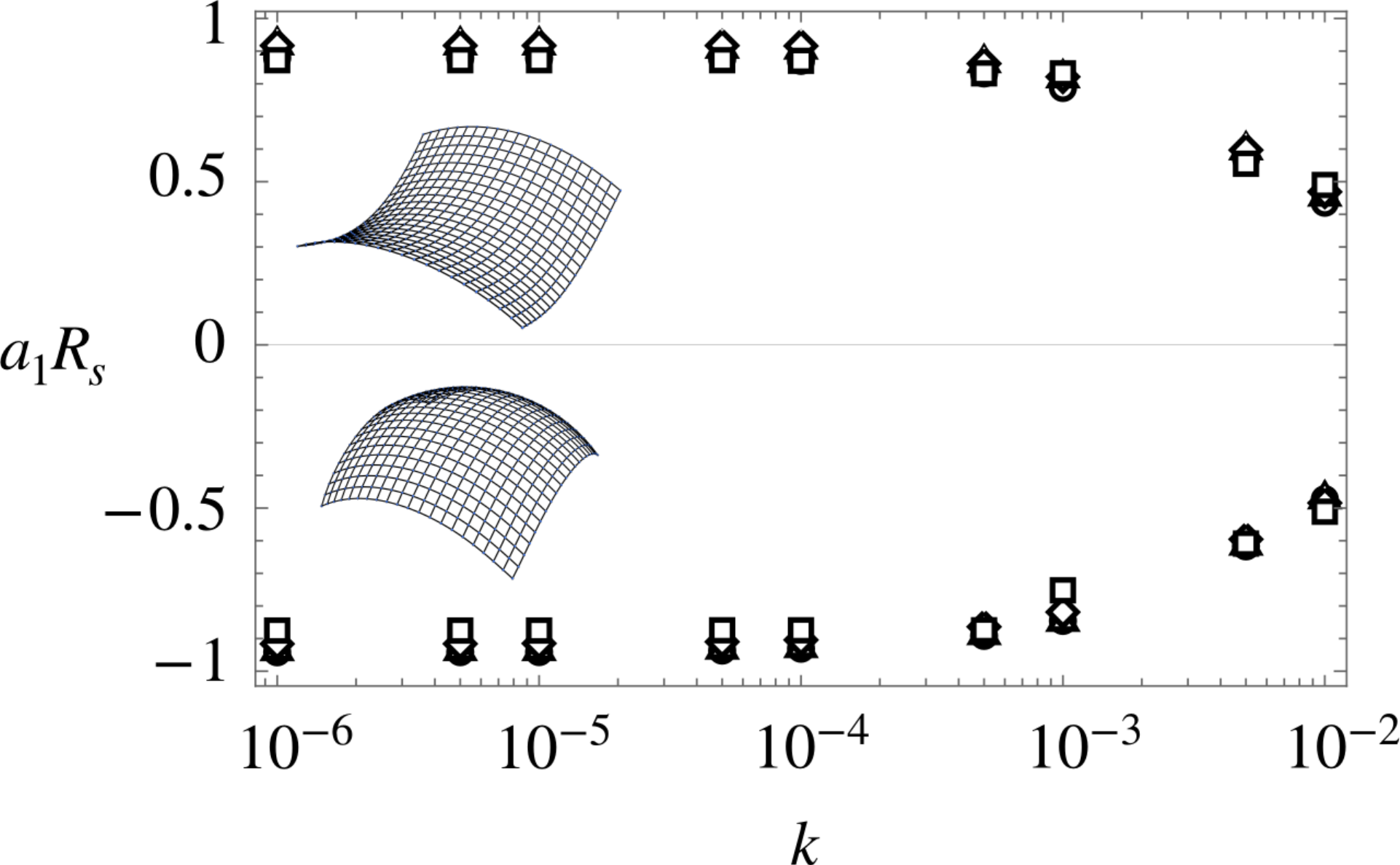}
    \caption{\label{fig7} Dependence of $a_1 R_s^2$ on the torsional modulus $k$ for $R_s = 6$, $10$, $20$ and $40$ with both a spherical and a saddle geometry.}
\end{figure}

In Fig. \ref{fig7}, we plot $a_1 R_s^2$ as a function of the torsional stiffness, $k$, for both positive and negative Gaussian curvatures, showing that the curves collapse for different curvatures and are consistent with the functional form expected from our theory.

\section{Conclusions}
In this paper, we have developed a continuum model for mechanical metamaterials confined to but able to slide on curved surfaces. Our approach has two steps. The first step introduces an auxiliary field, $A$, which is used to separate the soft mode from the stiffer deformation modes. In the second step, we rewrite the elastic energy in terms of $A$, gradients of $A$ and orthogonal strains $\gamma_{i j} - A G_{i j}$. We explicitly considered both dilational metamaterials, which have a soft deformation mode under isotropic dilation, and shear metamaterials, which are soft under simple shear. In both cases, the metamaterial, through its internal soft mode, can absorb some of the Gaussian curvature of the surface upon which it is draped and, consequently, screen the Gaussian curvature.
To our knowledge, this is the first generalized elasticity approach to metamaterial sheets on curved surfaces. Our approach is basically phenomenological in that we are unable to easily determine the relationship between microscopic parameters and the elastic moduli in the energy. It would be interesting to supplement this approach with a more detailed calculation in the vein of Ref. \cite{Czajkowski2022}.

In our theoretical analysis, we have neglected bending energy contributions and we find reasonable agreement between simulations and theory in this limit. It would be quite interesting to further explore the balance between bending and the soft deformations of the metamaterial sheets. We expect this will play an important role in geometrically-frustrated metamaterial sheets, in which bending and stretching are inherently incompatible.
The balance between stretching and bending in a traditional elastic sheet is controlled by the F\"oppl-von K\`arm\`an number, which dictates when stretching modes are expelled in the bulk of a sheet. In a metamaterial, one expects additional dimensionless numbers associated with the soft deformation that competes with the F\"oppl-von K\'arm\'an number. How this idea plays out in more complex geometries than flat domains remains unclear.

Finally, we note that our general approach can be extended readily in two ways. First, we can consider metamaterials with more than one soft deformation by extending $A$ to a multi-component field. Second, a metamaterial sheet with nonlinear soft modes can be represented through the function $W_0(A)$ in Eq. (\ref{eq:W}). A bistable unit cell, for which $W_0(A)$ would be quartic, might naturally exhibit domain walls and solitons. Thus, our approach opens the door to exploring how nonlinear deformations couple more broadly to elasticity in metamaterial sheets.
\section*{Conflicts of interest}
There are no conflicts to declare.
\section*{Acknowledgements}
We are thankful for useful conversations and several helpful comments by
Greg Grason and Michael Wang. We acknowledge funding from the National
Science Foundation through Grant No. NSF DMR-2217543.



\renewcommand\refname{References}

\bibliography{mechmat} 

\pagebreak
\widetext
\begin{center}
\textbf{\large Supplemental Material:\\ Curvature screening in draped mechanical metamaterial sheets}
\end{center}
\setcounter{equation}{0}
\setcounter{section}{0}
\setcounter{figure}{0}
\setcounter{table}{0}
\setcounter{page}{1}
\makeatletter

\renewcommand{\theequation}{S\arabic{equation}}
\renewcommand{\thefigure}{S\arabic{figure}}
\renewcommand{\bibnumfmt}[1]{[S#1]}
\renewcommand{\citenumfont}[1]{S#1}
\def\thepage{S\arabic{page}}

\section{Boundary layer analysis}

\subsection{Dilational metamaterial}
For dilational metamaterial sheets, the governing equations of motion in the  F\"{o}ppl-Von K\'{a}rm\'{a}n limit are 
\begin{eqnarray}\label{equation1}
                                \triangle (\chi/Y)    &=& (\kappa_0/Y) A - (\kappa_1/Y) \Delta A,\\
\label{equation2}               \triangle^2 (\chi/Y) &=& - G -\Delta A
\end{eqnarray}
where $\kappa_1/Y L^2 $ is assumed to be a small, dimensionless number, where L is the dimension of the sheet. Here $Y= 4 \mu$ as per the main text. And 
The boundary conditions are
\begin{eqnarray}
    \hat{\mathbf{n}} \cdot \nabla A|_{B} &=& 0,\\
    \sigma \cdot \hat{\mathbf{n}}|_{B} &=& 0,
\end{eqnarray}
where $\sigma$ is the stress tensor, $|_B$ denotes the restriction of the functions to the boundary, $\hat{\mathbf{n}}$ is the unit vector normal to the boundary but tangent to the metamaterial sheet, and repeated indices are summed.

The presence of a small parameter $\kappa_1/YL^2$ multiplying the highest derivative of $A$ suggests the solution will take the form of a slowly varying bulk term and a rapidly varying boundary layer. An approximate solution may be obtained as described in the next sections.

\subsubsection{Outside the boundary layer}
Since $\epsilon = \frac{\kappa_1}{YL^2}$ is small, we look for a solution in the bulk of the sheet of the form,
\begin{eqnarray}\label{perturbation}
    A &=& A_0 + \epsilon A_1+ O(\epsilon^2)\\
    \chi &=& \chi_0 + \epsilon \chi_1 + O(\epsilon^2).
\end{eqnarray}
Substituting into Eqs. (\ref{equation1}) and (\ref{equation2}), we obtain 
\begin{eqnarray}
    \triangle \chi_0 &=& \kappa_0 A_0\\
    \triangle A_0 &=&- \frac{Y}{Y+\kappa_0}G
\end{eqnarray}  
to lowest order in $\epsilon$.  Assuming axisymmetry and setting $Y=4 \mu$ we therefore obtain $a_2 =  - \frac{\mu G }{(4 \mu + \kappa_0)}$, thus giving
\begin{equation}
    A_0 = - \frac{\mu G}{4\mu+\kappa_0} r^2 + a_0,
\end{equation}
,where $a_0$ is a constant.
\subsubsection{Inside the boundary layer}
To obtain the solution near the boundary, assume the boundary layer has thickness $\delta$ and make the coordinate substitution $\eta =(R-r)/\delta$. From the equations of motion, we obtain the following effective equation in A
\begin{eqnarray}
    \frac{\kappa_1/4\mu}{\delta^2}\triangle^2 A_{in} = (1+ \frac{\kappa_0}{4\mu})\triangle A_{in}.
\end{eqnarray}
Setting the boundary layer thickness $\delta = \sqrt{\frac{\kappa_1}{4\mu+ \kappa_0}} = l_{sc}$, we obtain
\begin{equation}
    A(\eta)_{in} = c_1 (I_0(\eta)-1) + c_2 Y_0(-i \eta)
\end{equation}
\subsubsection{Full solution}
Imposing the boundary condition $A'(R) = 0$ on the boundary layer solution, in the limit $\delta \to 0$ that we can evaluate A

\begin{equation}
    A = a_0- \frac{4\mu G}{4 \mu+\kappa_0}\left(\frac{ r^2}{4} - l_{sc} \frac{ R}{2} \frac{I_0\left(\frac{r}{l_{sc}}\right)}{I_1\left(\frac{R}{l_{sc}}\right)}\right)
\end{equation}
The coefficient $a_0$ can be obtained by minimizing the energy for A we have obtained. 
Therefore, up to leading order in the screening length we have 
\begin{equation}
    A = \frac{\mu G (R^2 - 12 l_{sc}^2)}{3(\kappa_0 + 4 \mu)} - \frac{4\mu G}{4 \mu+\kappa_0}\left(\frac{ r^2}{4} - l_{sc} \frac{ R}{2} \frac{I_0\left(\frac{r}{l_{sc}}\right)}{I_1\left(\frac{R}{l_{sc}}\right)}\right)
\end{equation}
Similarly to ensure vanishing normal and shear stresses on the boundary, i.e, $\chi'(R) = 0$, we get $\chi$ up to appropriate order in $\delta$
\begin{equation}
    \chi = d_0  + \frac{G \mu \kappa_0 R^2 }{12 (\kappa_0 + 4 \mu)} r^2 -\frac{G \mu \kappa_0 }{16(\kappa_0 + 4\mu)} r^4  + l_{sc}^2\frac{\mu G}{4\mu +\kappa_0}\left(4 \mu r^2   - \frac{ R}{2} \frac{I_0\left(\frac{r}{l_{sc}}\right)}{I_1\left(\frac{R}{l_{sc}}\right)}\right)
\end{equation}
Note that the harmonic function in $\chi$ in the main text with the choice of boundary conditions reduces to the constant $d_0$. 

\subsection{Simple shear metamaterial}
For simple shear sheets, in Cartesian coordinates, the equilibrium equations are 
\begin{eqnarray}\label{equation1shear}
                        \partial_1\partial_2\chi   &=& -\kappa_0 A + \kappa_1 \triangle A\\
\label{equation2shear}  \frac{1}{{Y}} \triangle^2 \chi-\frac{1}{\mu}\partial_1^2\partial_2^2\chi &=& - G +\partial_1\partial_2 A
\end{eqnarray}

\subsubsection{Outside the boundary layer}
Outside the boundary layer,
\begin{eqnarray}\label{perturbationshear}
    A &=& A_0 + \epsilon A_1+ O(\epsilon^2),\\
    \chi &=& \chi_0 + \epsilon \chi_1 + O(\epsilon^2)
\end{eqnarray}
and so
\begin{eqnarray}
    \partial_1\partial_2\chi_0   &=& -\kappa_0 A_0,\\
    \frac{1}{{Y}} \triangle^2 \chi_0-\frac{1}{\mu}\partial_1^2\partial_2^2\chi_0 &=& - G +\partial_1\partial_2 A_0.    
\end{eqnarray}

We obtain the following working equation in $\chi_0$,
    \begin{equation}
        \frac{1}{Y}\left(\partial_1^4 + \partial_2^4 \right)\chi_0 + \left(\frac{1}{\kappa_0}+2\frac{1}{Y}- \frac{1}{\mu}\right)\partial_1^2\partial_2^2\chi_0 = -G.
    \end{equation}
We can guess the form of the simplest solution for the limit $\kappa_1 \to 0$ and $\kappa_0 \ll Y$ to be
\begin{eqnarray}
      \chi_0 &=& \frac{-\kappa_0~\mu~Y  G/4}{\mu Y + 2\kappa_0 \mu -\kappa_0 Y} x^2y^2 + P(x) + Q(y)\\
      A_0 &=& \frac{\mu~Y  G}{\mu Y + 2\kappa_0 \mu -\kappa_0 Y} xy
\end{eqnarray}
These functions do not satisfy the boundary conditions and are corrected by an additional exponentially growing term inside the boundary layer whose thickness scales with $\kappa_1$.

\subsubsection{Inside the boundary layer}
The working equations are the following
 \begin{eqnarray}\label{blequation1}
    \partial_1\partial_2\chi   &=& -\kappa_0 A + \kappa_1 \triangle A\\
              \label{blequantion2} \frac{1}{{Y}} \triangle^2 \chi-\frac{1}{\mu}\partial_1^2\partial_2^2\chi &=& - G +\partial_1\partial_2 A
\end{eqnarray}
Near boundary $x=L$, the $y$ derivatives are small and $x$ derivatives are large. Introducing an internal variable, $\eta = (L-x)/\delta$, the equations become,
\begin{eqnarray}
    \label{boundarylayerequations}
    -\delta\partial_{\eta}\partial_2\chi   &=& -\delta^2\kappa_0 A + \kappa_1  \left(\partial_{\eta}^2 + \delta^2\partial_2^2\right) A\\
              \label{boundarylayerequations2}  -\delta^3\partial_{\eta}\partial_2 A &=& \frac{1}{{Y}} \left(\partial_{\eta}^4 + \delta^4\partial_2^4+ 2\delta^2\partial_{\eta}^2\partial_2^2\right) \chi \\
              &-&\frac{1}{\mu}\delta^2\partial_{\eta}^2\partial_2^2\chi 
\end{eqnarray}
If the $\chi$ and $A$ are sufficiently slowly varying in $y$ near $x=L$, then
\begin{eqnarray}
    \delta^2 \kappa_0/\kappa_1 A &=& \partial_{\eta}^2 A \\
    - Y \delta^3 \partial_{\eta}\partial_2 A &=& \partial_{\eta}^4 \chi 
\end{eqnarray}
the solutions are of the following form 
\begin{eqnarray}
    \label{Aexp}
    A(x,y) &=&  \frac{\mu~Y  G}{\mu Y + 2\kappa_0 \mu -\kappa_0 Y} xy + f(y)~\left(e^{-(L-x)/l_{shear}}-e^{-(L+x)/l_{shear}}\right)\nonumber \\
    &+& g(x)~\left(e^{-(L-y)/l_{shear}}-e^{-(L+y)/l_{shear}}\right)
\end{eqnarray}
where $\delta = l_{shear}= \sqrt{\kappa_1/\kappa_0}$.
Correspondingly, $\chi$ is
\begin{eqnarray}
\label{chiexp}
 \chi(x,y) &=&  \frac{-\kappa_0~\mu~Y  G/4}{\mu Y + 2\kappa_0 \mu -\kappa_0 Y} x^2y^2 ~ + p(x^2+y^2)\nonumber \\&-&Y l_{shear}^3 \Bigg(f'(y)~\left(e^{-(L-x)/l_{shear}}+e^{-(L+x)/l_{shear}}\right)\nonumber\\ &~ & ~~~~~+g'(x)~\left(e^{-(L-y)/l_{shear}}+e^{-(L+y)/l_{shear}}\right)\Bigg) 
\end{eqnarray}
For \ref{Aexp} and \ref{chiexp}, to satisfy \ref{blequation1} and \ref{blequation2}, the forms of $f(x)$ and $g(y)$ are
\begin{eqnarray*}
    f(y) &=& a_1 y\\
    g(x) &=& a_1 x
\end{eqnarray*}
We determine the constants $a_1$ and $p$ by satisfying the boundary condition in the limit $\kappa_1 \rightarrow 0$ and $\kappa_0 G \rightarrow 0$. Thus,
\begin{eqnarray*}
    a_1 &=& \frac{\mu  Y G l_{shear}\left(1+\tanh \left(L \sqrt{\frac{\kappa_0}{\kappa_1}}\right)\right)}{2 ( \mu Y +2 \kappa_0\mu -\kappa_0 Y)}\\
    p &=&  \frac{\kappa_0~\mu~Y  G}{4(\mu Y + 2\kappa_0 \mu -\kappa_0 Y)}L^2
\end{eqnarray*}
Hence, the fields simplify to the following final expressions
\begin{eqnarray*}
    A(x,y) &=& \frac{\mu  Y G }{  \mu Y +2 \kappa_0\mu -\kappa_0 Y}\left( x y + l_{shear} \frac{ x~ \sinh(y/l_{shear} )+ y ~\sinh(x/l_{shear})}{\cosh(L/l_{shear})} \right)\\
    \chi(x,y) &=& \frac{\mu Y G/4}{\mu Y+ 2\kappa_0 \mu -\kappa_0 Y}\Big(L^2(x^2+y^2) -x^2y^2 \\&~& ~~~~~- 4 Y l_{shear}^4~ \frac{\cosh(x/l_{shear})+\cosh(y/l_{shear})}{\cosh \left(L/l_{shear}\right)} \Big)
\end{eqnarray*}
where, we have small parameters $\kappa_1$ and $\kappa_0$ satisfying $ l_{shear} \ll L$ and $G L^2\kappa_0/Y \rightarrow0 $. Thus $l_{shear} \ll \sqrt{Y/G\kappa_0}$, which gives $\kappa_1/Y \ll 1/G $ or $G \ll Y/\kappa_1$. Hence, this solution is valid in this limit of sufficiently weakly curved substrate.\\

\section{Numerics}
\begin{figure}[ht]
    \includegraphics[width=0.8\textwidth]{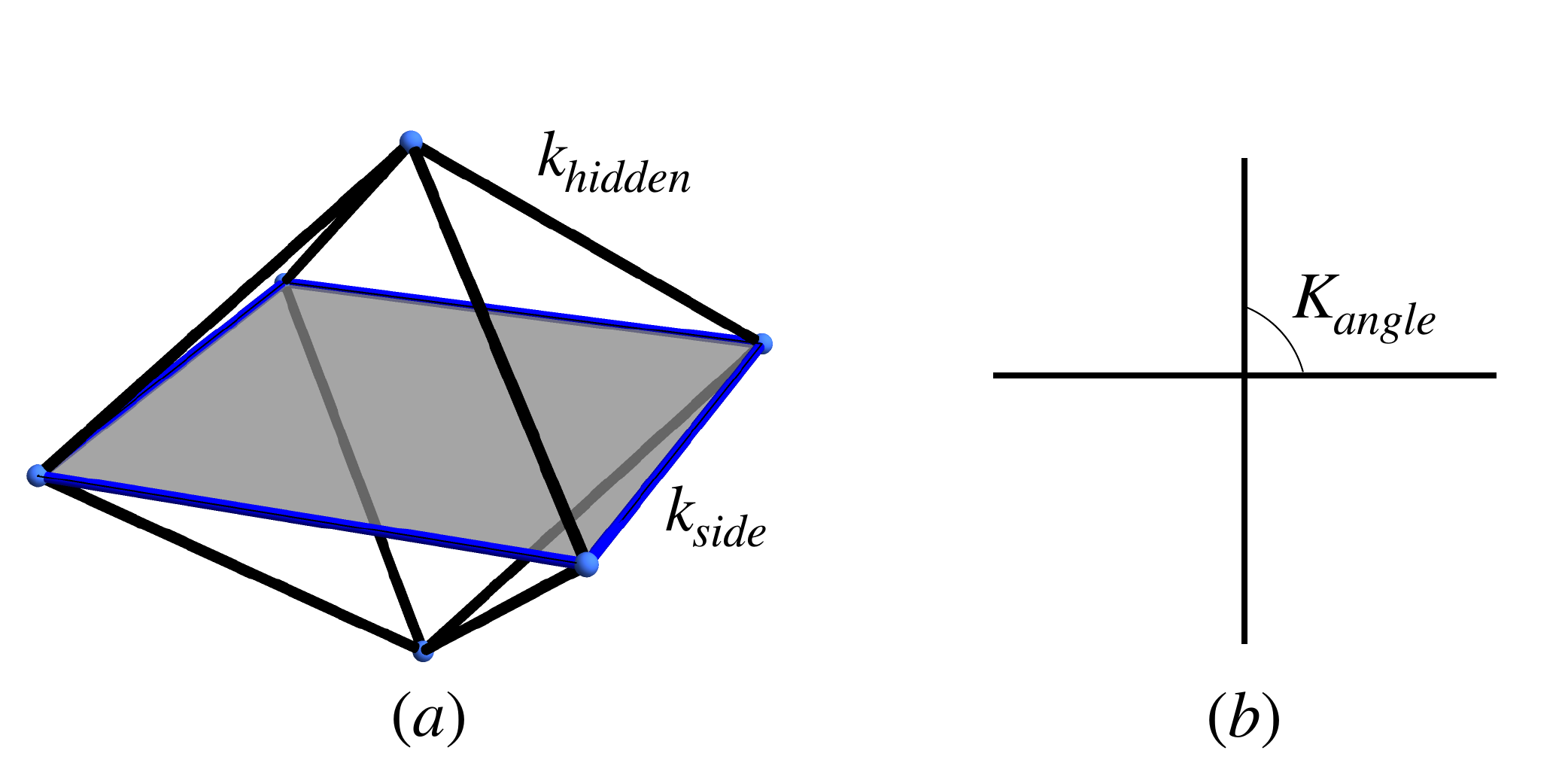}
    \caption{\label{fig-block} (a) One unit of the dilational metamaterial. (b) A vertex of the shear metamaterial.}
\end{figure}
The metamaterials are modeled as linear and torsional spring networks with each spring having energy
\begin{equation}
    E_{lin} = \frac{k}{2} \left( \frac{l^2 - \bar{l}^2}{2 \bar{l}^2} \right)^2,
\end{equation}
where $l$ is the spring length, $\bar{l}$ the equilibrium length, and $k$ the spring constant. For small strains, $l \approx \bar{l} + \delta l$ and the energy reduces to
\begin{equation}
    E_{lin} \approx \frac{k}{2} \left( \frac{l}{\bar{l}} \right)
\end{equation}

Torsional springs are modeled as
\begin{equation}
    E_{tor} = \frac{K}{2} \left( \theta - \bar{\theta} \right)^2,
\end{equation}
where $\theta$ is the angle between a pair of edges and $\bar{\theta}$ is their equilibrium angle, and $K$ is the torsional spring constant.

Finally, each geometrically-confined vertex is modeled with energy
\begin{equation}
    E_{con} = \frac{k_{con}}{2} \left( z - h(x,y) \right)^2,
\end{equation}
where $h(x,y)$ is the height function of a surface and $(x,y,z)$ are the coordinates of the vertex. The constraint $k_{con}$ is set to 100 times the largest modulus and we have confirmed that $k_{con}$ is always sufficiently large so as to not change the numerical results.

The total energy was minimized with the standard L-BFGS algorithm and conjugate gradient algorithm as implemented by Mathematica version 12.

\subsection{Dilational metamaterial}
The dilational metamaterial is constructed as a system of counter-rotating squares. Each square is actually modeled according to Fig. \ref{fig-block}, with hidden edges leading to a vertex above and below the centroid of the square. The sides of the square have spring constant $k_{side} = 1$ and $k_{hidden} = 10^{-3}$ sets the bending rigidity of the square. Torsional springs are placed at the vertices where squares meet.

\subsection{Shear metamaterial}
The shear metamaterial is built from a square lattice of springs with spring constant $k=1$. There is a torsional spring with spring constant $K_{bend}$ preventing the bending of each rod through the vertex and an additional torsional spring with spring constant $K_{angle}$ at one corner of each square of the metamaterial.
\end{document}